\documentclass[conference]{IEEEtran}

\usepackage{bm}
\usepackage{mathrsfs}
\usepackage{caption}
\usepackage{amsmath}
\allowdisplaybreaks[4]
\usepackage{lipsum}
\usepackage{mathtools}
\usepackage{subfigure}
\usepackage{amsmath} 
\usepackage{amssymb}
\usepackage{amsfonts}
\usepackage{verbatim} 
\usepackage{bm,color,soul}
\usepackage{algorithm}
\usepackage{algorithmic} 
\usepackage{cite}
\usepackage{caption}
\usepackage{stfloats}

\makeatletter
\renewcommand{\maketag@@@}[1]{\hbox{\m@th\normalsize\normalfont#1}}%
\makeatother
\newtheorem{remark}{Remark}
\usepackage[colorlinks=black, linkcolor=black, anchorcolor=black, citecolor=black]{hyperref} 
\hypersetup{hidelinks}

\usepackage{makecell}

\ifCLASSINFOpdf
\else
\fi
\usepackage{array}
\usepackage{balance}

\IEEEoverridecommandlockouts

\begin{document}
	%
	\title{Sensing-assisted Near-field Energy Beam Focusing with ELAA Over Non-stationary Channels}
	\author{
		\IEEEauthorblockN{Li Zhang$^{1}$, Zixiang Ren$^{1,2}$, Yuan Fang$^{2,3}$, Ling Qiu$^{1}$, and Jie Xu$^{4,2}$}
		\IEEEauthorblockA{$^1$Key Laboratory of Wireless-Optical Communications, Chinese Academy of Sciences,  \\
			School of Information Science and Technology,
			University of Science and Technology of China}
		\IEEEauthorblockA{$^2$Shenzhen Future Network of Intelligence Institute, The Chinese University of Hong Kong (Shenzhen)}  
		\IEEEauthorblockA{$^3$Department of Electrical Engineering, City University of Hong Kong}
		\IEEEauthorblockA{$^4$School of Science and Engineering, The Chinese University of Hong Kong (Shenzhen)}
		\IEEEauthorblockA
		{E-mails: {lzhang0228}@mail.ustc.edu.cn,
			{fangyuan}@cuhk.edu.cn, {rzx66}@mail.ustc.edu.cn,\\	{lqiu}@ustc.edu.cn, {xujie}@cuhk.edu.cn}  \vspace{-0.8cm}  
		\thanks{Ling Qiu and Jie Xu are the corresponding authors.} 
	}	
	\maketitle
	\begin{abstract}
		This paper studies a novel training-free energy beam focusing approach for a near-field wireless power transfer (WPT) system with extremely large-scale antenna array (ELAA). In particular, we focus on the setup with one access point (AP) equipped with an extremely large-scale uniform planar array (UPA) serving multiple single-antenna energy receivers (ERs), in which the line-of-sight (LoS) dominated wireless channels are dependent on the relative positions of ERs and exhibit spatial non-stationarity.
		Different from conventional designs relying on training and feedback, we present a novel energy beam focusing design assisted by wireless radar sensing based on a two-stage transmission protocol. In the first stage, the AP performs wireless radar sensing to identify the ERs' visibility regions (VRs) and estimate their
		three-dimension (3D) positions for constructing the
		corresponding channel state information (CSI). In the second stage, the AP implements the
		transmit energy beam focusing based on the constructed CSI to
		efficiently charge these ERs. Under this setup, we first minimize the sensing duration in the first stage, while guaranteeing a specific accuracy threshold for position estimation.  Next, we optimize the energy beamformers at the AP in the second stage to maximize the weighted harvested
		energy among all ERs subject to the maximum transmit power constraint. In this approach, the time resource allocation between the two stages is properly designed to optimize the ultimate energy transfer performance. Numerical results show that the proposed design performs close to the performance upper bound with perfect VR and CSI and significantly outperforms other benchmark schemes. 
	\end{abstract}
	\vspace{-0.15cm}	
	\section{Introduction}
	\vspace{-0.1cm}	
	Future sixth-generation (6G) wireless networks are envisioned to support the connectivity of massive Internet-of-things (IoT) devices to realize the vision of connected everything and connected intelligence \cite{1}. However, many IoT devices are with small size and have limited power supply. Therefore, there is a growing need to find convenient and sustainable techniques for charging these devices. Wireless power transfer (WPT) has been considered a viable solution to charge a large number of IoT devices concurrently, in which base stations (BSs) or access points (APs) utilize radio frequency (RF) signals to wirelessly charge these devices as energy receives (ERs) \cite{2}.
	
	On the other hand, extremely large-scale antenna array (ELAA) is becoming an enabling technology in 6G to provide enhanced  wireless performance via exploiting the large array, multiplexing, and diversity gains, particularly at higher frequency bands such as millimeter wave (mmWave) and terahertz (THz). With ELAA, devices situated within the so-called Rayleigh distances are regarded as operating in the near-field region. Unlike the far-field region in which the electromagnetic fields can be approximated as plane waves, the near-field electromagnetic field takes the form of spherical wavefront. In particular, the BS/AP transmitter can form highly focused beams in both angular and distance domains for communications and WPT, thus mitigating interference in multi-user communication \cite{4} and significantly enhancing the energy transfer efficiency \cite{11,12}.
	
	The performance gain of energy beam focusing in ELAA heavily relies on the accuracy of channel state information (CSI) acquisition \cite{6}. However, traditional pilot-based channel estimation methods are highly costly due to the large dimension caused by ELAA. In practical high-frequency ELAA systems, obstacles in the environment may completely block the signals, leading to sparse multi-path channel components. Therefore, various prior works on channel estimation have exploited the inherent sparsity to reduce the pilot overhead in such systems. For example, for the case of near-field channel estimation, the authors in \cite{8} proposed to exploit the sparsity in the polar domain by sampling both the angular and distance ranges to construct a two-dimensional dictionary, based on which the corresponding compressed sensing (CS) algorithm is devised to estimate the channel. However, such methods  have a high storage burden and computational complexity. In addition, due to signal blockage and spherical wavefront propagation, the large-dimension channels also have the so-called spatial non-stationarity along the array space. In this case, the energy of each user is concentrated only on a part of the arrays, characterized by visibility region (VR). The VR makes the channel estimation problem even more difficult. The authors in \cite{9} utilized the received pilot sequence to identify the user's VR and estimate the channel within the obtained VR via the least squares (LS) method. However, the above channel estimation designs require the user receivers to perform active signal feedback transmission, which may not work well for WPT systems, as the active signal transmission would reduce the net energy harvested by the ERs.
	
	Integrating sensing as a new  functionality has been recognized as another key technology for 6G to enable integrated sensing and communication (ISAC) \cite{18} or even integrated sensing, communication, and powering (ISCAP) \cite{10,17}. In particular, the sensing function can be exploited to provide additional environmental information to facilitate the CSI acquisition for assisting both communication and WPT. While there have been various prior works investigating sensing-assisted communications under various different setups \cite{19,21}, the study on sensing-assisted WPT is still at its infancy. In the literature, our prior work \cite{13} studied energy beamforming based on constructed CSI by assuming the far-filed line-of-sight (LoS) channels, in which the AP performs radar sensing to estimate path gain and angle parameters of the ERs for constructing CSI in a two-dimensional (2D) plane. However, in near-field scenarios, the ERs are more likely to be situated in three-dimensional (3D) space, rendering these designs inapplicable for near-field WPT. In particular, due to the dominance of the LoS channel, the near-field channels are dependent on the relative positions between the AP and ERs, and are subject to the spatial non-stationarity with different VR towards each ER. Therefore, how to exploit radar sensing to achieve efficient  near-field energy beam focusing is an interesting topic that has not been investigated yet, thus motivating the current work. 
	
	This paper studies a sensing-assisted energy beam focusing approach in a near-field ELAA WPT system, in which a AP equipped with an extremely large-scale uniform planar array (UPA) wirelessly charges multiple single-antenna ERs. In particular, we present a two-stage transmission protocol for sensing-assisted near-field WPT, in which the whole transmission period is divided into two stages for radar sensing and energy transmission, respectively, and the first stage is further divided into $K$ slots with equal durations each for the radar sensing with one ER. In each slot of the first radar sensing stage, the AP sends radar sensing signals and collects the echo signals from one ER to identify the VR and estimate its three-dimension position. Then, the AP constructs the CSI of ERs based on the identified VR and estimated positions. In the second energy transmission stage, the AP performs energy beam focusing based on constructed CSI. Under this setup, we first minimize the sensing duration in the first stage based on the ERs’ estimated parameters from the previous block, while guaranteeing a given accuracy threshold for position estimation. Next, we optimize the energy beamformers at the AP in the second stage to maximize the weighted harvested
	energy among all ERs subject to the maximum transmit power constraint. In the proposed approach, the allocation of time resource between the two stages is properly designed to optimize the overall energy harvesting performance. Finally, numerical results show that the proposed design achieves performance close to the upper bound with perfect VR and CSI and superior performance compared to benchmark schemes. 
	
	\emph{Notations:} Matrices are denoted by bold uppercase letters, and vectors are represented by bold lowercase letters. For a square matrix $\mathbf{A}$, $\operatorname{tr}\left(\mathbf{A}\right)$ denotes
	its trace, and $\mathbf{A} \succeq \mathbf{0}$ means that $\mathbf{A}$ is positive semi-definite. For a vector $\mathbf{b}$, $\operatorname{diag}(\mathbf{b})$ denotes a diagonal matrix with $\mathbf{b}$ being its diagonal elements. For an arbitrary-size matrix $\mathbf{b}$, $\operatorname{rank}\left(\mathbf{b}\right)$, $\mathbf{b}^{T}$, $\mathbf{b}^{H}$, and $\mathbf{b}^{*}$ denote
	its rank, transpose, conjugate transpose, and 
	conjugate, respectively. $\mathbb{E}(\cdot)$
	represents the stochastic expectation, $|\cdot|$ denotes the absolute value of a scalar, and $\|\cdot\|$ denotes the
	Euclidean norm of a vector. $\left[\cdot\right]_{n}$ denotes the $n\text{-th}$ entry of a vector, and $\mathbb{C}^{M \times N}$ denotes
	the space of $M \times N$ complex matrices. $\mathbf{A}\odot \mathbf{B}$ represents the Hadamard
	product of two matrices $\mathbf{A}$ and $\mathbf{B}$. Furthermore, we denote $j=\sqrt{-1}$.
	\vspace{-0.1cm}
	\section{System Model}
	\begin{figure}
		\setlength{\abovecaptionskip}{-16pt}
		\setlength{\belowcaptionskip}{-16pt}
		\centering
			\centering
			\includegraphics[width= 0.38\textwidth]{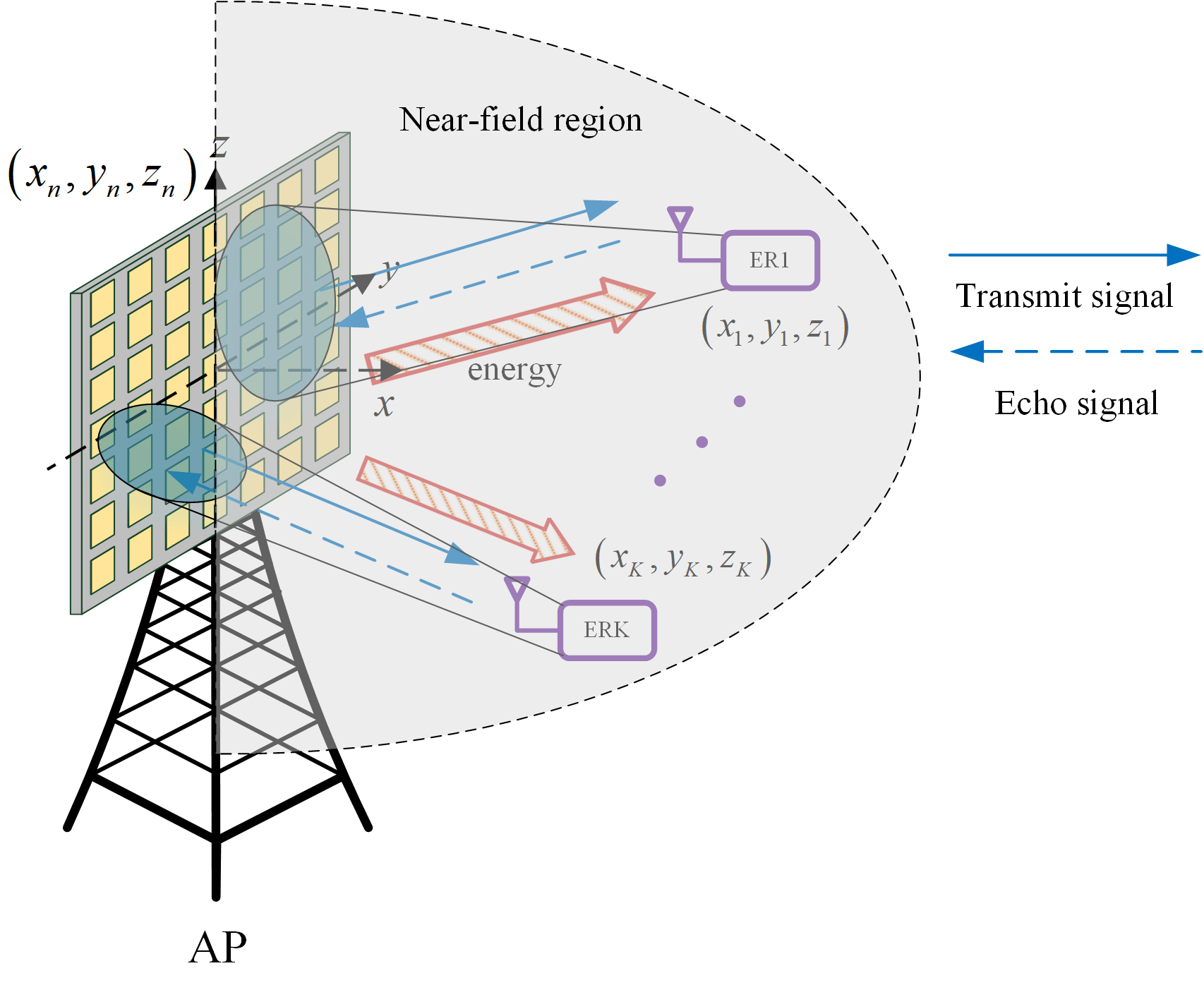}
		\DeclareGraphicsExtensions.
		\captionsetup{font={scriptsize},skip=3pt}
		\caption{Illustration of a near-field sensing-assisted WPT system.}
		\label{figure1}
	\end{figure}	
	We consider a near-field ELAA WPT system as illustrated in Fig. 1, in which an AP equipped with an extremely large-scale UPA transmits  energy to $K$ single-antenna ERs. We consider a
	mono-static radar sensing setup at the AP, in which the transmitter and sensing
	receiver are co-located.  Let $\mathcal{K} \triangleq\{1, \ldots, K\}$ denote the set of ERs. It is assumed that the UPA in a rectangular shape with a total of $N = N_{y} \times N_{z}$ antennas, where $N_{y}$ and $N_{z}$ are the numbers of antennas along the y-axis and the z-axis, and each ER is equipped with a RF energy harvesting module and a backscatter modulation module.
	
	As shown in Fig. 2, we consider the block-based transmission. Let $T$ denote the duration of each transmission block in the number of symbols. It is assumed that the locations of ERs and their wireless channels remain constant within each block but may vary across different blocks due to the mobility of ERs. 	
	\begin{figure}
		\setlength{\abovecaptionskip}{-16pt}
		\setlength{\belowcaptionskip}{-16pt}
		\centering
			\centering
			\includegraphics[width= 0.32\textwidth]{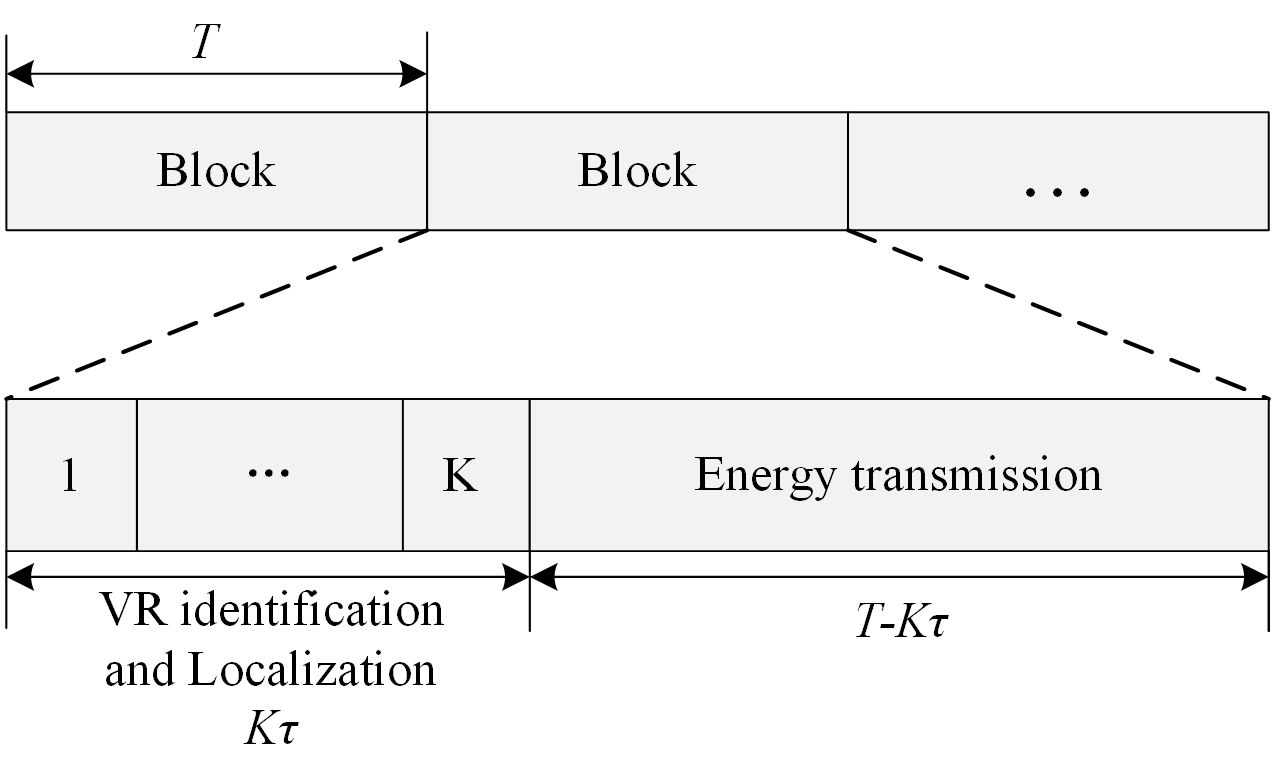}
		\DeclareGraphicsExtensions.
		\captionsetup{font={scriptsize},skip=3pt}
		\caption{Block-based transmission with two-stage protocol.}
		\label{figure2}
	\end{figure}	
	We consider a two-stage transmission protocol, in which the transmission block with duration $T$ is divided into two stages with $K\tau$ symbols for VR identification and localization and $T-K\tau$ symbols for energy transmission, respectively. In addition, the duration $K\tau$ in the first stage is segmented into $K$ slots. For practical implementation, we assume an identical duration $\tau$ for each slot. Here, the duration $\tau$ is a decision variable to be determined based on the estimation results in the previous block. In each slot of the first stage, the AP transmits a specified waveform to identify the VR and estimate the position of one ER in the current block for constructing the corresponding CSI. In the second stage, the AP performs the transmit energy beam focusing based on the constructed CSI in the first stage of the current block to efficiently charge multiple ERs.
	
	Suppose that the wireless channels are dominated by LoS paths and are non-stationary. Accordingly, we consider the near-field LoS channel model by taking into account the VR. Let the center of the UPA be the origin of the yoz plane. Let $\mathbf{l}^{t}_{n}=\left(x_{n}^{t}, y_{n}^{t}, z_{n}^{t}\right)$, $n \in \{1, 2, \ldots, N\}$, represent the location of the $n \text{-th}$ antenna at the AP, and $\mathbf{l}_{k}=\left(x_{k}, y_{k}, z_{k}\right)$ denote the location of ER $k$. In this case, the steering vector at the AP with respect to ER $k$ is given by \cite{14}	
	\begin{equation}\label{steering vector}			
		\begin{aligned}
			\setlength\abovedisplayskip{4pt}		
			\setlength\belowdisplayskip{4pt}
			\mathbf{a}\left(\mathbf{l}_{k}\right)=\left[a_{1}\left(\mathbf{l}_{k}\right) e^{-\mathrm{j} \frac{2 \pi}{\lambda}\left\|\mathbf{l}^{t}_{1}-\mathbf{l}_{k}\right\|}, \ldots, a_{N}\left(\mathbf{l}_{k}\right) e^{-\mathrm{j} \frac{2 \pi}{\lambda}\left\|\mathbf{l}^{t}_{N}-\mathbf{l}_{k}\right\|}\right]^{T},
		\end{aligned}
	\end{equation}
	where $\lambda$ denotes the carrier wavelength, and $a_{n}\left(\mathbf{l}_{k}\right)=\frac{\lambda}{4 \pi\left\|\mathbf{l}_{k}-\mathbf{l}^{t}_{n}\right\|}$ denotes the distance-dependent channel amplitude from the $n \text{-th}$ antenna to ER $k$ based on the free space path-loss model. Note that the phase and amplitude in the array steering vector $\mathbf{a}\left(\mathbf{l}_{k}\right)$ in \eqref{steering vector} depend on the relative positions between the antenna elements and ER $k$. Therefore, the model in \eqref{steering vector} takes into account the precise spherical wavefront with variations in both channel phases and amplitudes across different antenna elements.  
	
	To capture the channel non-stationarity, we denote the VR of ER $k$ at the AP as $\boldsymbol{\Phi}_{k}$, which comprises the antenna elements that are not blocked. For simplicity, the VR is represented as a contiguous sub-array of the UPA, as commonly adopted in the literature \cite{15}.\footnote{In practice, the VR induced by blockage may not be contiguous. Identifying a non-contiguous VR will be addressed in future work.} It is assumed that the size of the VR is larger than $\eta N$, where $0 < \eta < 1$ is a constant scaling factor. By assuming the starting and ending antenna indices in $\boldsymbol{\Phi}_{k}$ as $\bar{n}_{k}$ and $\underline{n}_{k}$, we have
	\begin{equation}	
		\setlength\abovedisplayskip{4pt}		
		\setlength\belowdisplayskip{4pt}	
		\boldsymbol{\Phi}_{k}=\left\{\bar{n}_{k}, \bar{n}_{k}+1, \ldots, \underline{n}_{k} - 1, \underline{n}_{k}\right\},
	\end{equation}
	where $1 \leq \bar{n}_{k} < \underline{n}_{k} \leq N$ and $\underline{n}_{k} - \bar{n}_{k} \geq \eta N$.
	Accordingly, we define the VR cover vector of ER $k$ as $\mathbf{g}\left(\boldsymbol{\Phi}_{k}\right) \in \mathbb{R}^{N \times 1}$, whose $n \text{-th}$ element is given by
	\begin{equation}	
		\setlength\abovedisplayskip{4pt}		
		\setlength\belowdisplayskip{4pt}	
		[\mathbf{g}(\boldsymbol{\Phi}_{k})]_{n}=\left\{\begin{array}{ll}
			1, & \text { if } n \in \boldsymbol{\Phi}_{k}, \\
			0, & \text { otherwise}.
		\end{array}\right.
	\end{equation}
	Accordingly, by assuming that each antenna element is omni-directional with unit antenna gain, the spatial non-stationary channel vector between AP and ER $k$ is given by
	\begin{equation}	
		\setlength\abovedisplayskip{4pt}		
		\setlength\belowdisplayskip{4pt}	
		\mathbf{h}_{k}=\mathbf{a}\left(\mathbf{l}_{k}\right) \odot \mathbf{g}\left(\boldsymbol{\Phi}_{k}\right).
	\end{equation}
	\subsection{Radar Sensing Stage}
	First, we consider the VR identification and localization in the first stage. Let $\mathbf{x}(t)=\sqrt{\frac{P_{\text{max}}}{N}} [1, \ldots, 1] \in \mathbb{C}^{N \times 1}$ denote the specified transmitted signal at symbol $t\in\{1,\ldots, K\tau\}$. We consider that ERs adopt on/off keying (OOK) to modulate received signal. Let $c_{k}(t)$ denote the modulation signal of ER $k$ at symbol $t$, i.e., $c_{k}(t)\in \left\{0,1\right\}$. In each slot $k$ of the first stage, only ER $k$ backscatters received signal by switching its load impedance. Thus, the received echo signal by the AP from ER $k$ is given by
	\begin{equation}\label{y(k)}				
		\begin{aligned}
			\setlength\abovedisplayskip{4pt}		
			\setlength\belowdisplayskip{4pt}
			\mathbf{y}_{k}(t)=b_{k} \mathbf{h}_{k} \mathbf{h}_{k}  ^{T} \mathbf{x}(t) + \mathbf{z}(t), t \in \tilde{\mathcal T},
		\end{aligned}
	\end{equation}
	where $b_{k} \in \mathbb{C}$ represents the target complex
	reflection coefficients proportional to the radar cross section
	(RCS) of ER $k$, $\mathbf{z}(t)$ denotes the additive white Gaussian noise (AWGN) with mean zero and
	covariance $\sigma_{r}^{2}\mathbf{I}$, and $\tilde{\mathcal T}=\{(k-1)\tau+1,\ldots, k\tau\}$.
	
	Note that the received signal $\mathbf{y}_{k}(t)$ in \eqref{y(k)} captures the inherent structure of the VR. Specifically, the average power of the entries is higher within the VR than that outside the VR. Thus, we employ a sliding window-based method to identify the VR of ER $k$. Let $\bar{\mathbf{y}}_{k} \in \mathbb{C}^{N \times 1}$ denote the summation of received signal $\mathbf{y}_{k}(t)$ from the ER $k$ during $t \in \tilde{\mathcal T}$, and thus we have
		\begin{equation}\label{tilde y}				
		\begin{aligned}
			\setlength\abovedisplayskip{4pt}		
			\setlength\belowdisplayskip{4pt}
			\bar{\mathbf{y}}_{k}=\sum_{t=(k-1)\tau+1}^{k\tau} b_{k} \mathbf{h}_{k} \mathbf{h}_{k}  ^{T} \mathbf{x}(t) + \mathbf{z}(t). 
		\end{aligned}
	\end{equation}
	Based on \eqref{tilde y}, we aim to identify a region that is as compact as possible while encompassing the primary power of $\bar{\mathbf{y}}_{k}$ as the VR estimate \cite{15}. A window with boundaries $\left[\bar{n}, \underline{n}\right]$ is utilized to represent VR, and this window is adjusted by modifying $\bar{n}$ and $\underline{n}$ to find the VR. In particular, we  maximize the received power within the window via maximizing $\sum_{n=\bar{n}}^{\underline{n}}|\left[\bar{\mathbf{y}}_{k}\right]_{n}|$, which is also equivalent to minimizing $\sum_{n=1}^{\bar{n}-1}|\left[\bar{\mathbf{y}}_{k}\right]_{n}|+\sum_{n=\underline{n}+1}^{N}|\left[\bar{\mathbf{y}}_{k}\right]_{n}|$. At the same time, we need to minimize the size of window, i.e., $\underline{n}-\bar{n}+1$. We define a function as
	\begin{align}	
		\setlength\abovedisplayskip{4pt}
		\setlength\belowdisplayskip{4pt}	
		f_{k}(\bar{n}, \underline{n})=\sum_{n=1}^{\bar{n}-1}|\left[\bar{\mathbf{y}}_{k}\right]_{n}|+\sum_{n=\underline{n}+1}^{N}|\left[\bar{\mathbf{y}}_{k}\right]_{n}|+\alpha(\underline{n}-\bar{n}+1),
	\end{align}
	where $\alpha$ is a scaling factor. The estimate of VR is obtained by finding $\left(\bar{n}, \underline{n}\right)$ that minimizes the function $f_{k}(\bar{n}, \underline{n})$:
	\begin{subequations}		
		\begin{align}
			\setlength\abovedisplayskip{4pt}		
			\setlength\belowdisplayskip{4pt}
			\left(\hat{\bar{n}}_{k}, \underline{\hat{n}}_{k}\right) & =\arg \min _{(\bar{n}, \underline{n})} f_{k}(\bar{n}, \underline{n}) \\
			\text { s.t. } \quad \bar{n} & =1, \ldots,(1-\eta) N ,\\
			\underline{n} & =\bar{n}+\eta N, \ldots, N.
		\end{align}
	\end{subequations}
	The optimal solution is achieved when $	f_{k}(\bar{n}, \underline{n})$ is minimized with $\bar{n}=\bar{n}_{k}$ and $\underline{n}=\underline{n}_{k}$, i.e., the window precisely covers the entire VR. The pre-defined scaling factor $\alpha$  should be appropriately set and satisfy $P_{\text{out}} < \alpha < P_{\text{in}}$ to accurately obtain $\bar{n}_{k}$ and $\underline{n}_{k}$, where $P_{\text{out}}$ and $P_{\text{in}}$ represent the average values of $|\left[\bar{\mathbf{y}}_{k}\right]_{n}|$ outside and inside the VR, respectively. In practice, $P_{\text{out}}$ is influenced by additive Gaussian noise, and the value of $P_{\text{in}}$ is primarily determined by the distance between the AP and the ERs. However, it is difficult to estimate $P_{\text{out}}$ and $P_{\text{in}}$ without knowledge of the VR. To address this problem, we propose a scheme as follows. First, based on the fact that $P_{\text{out}}<P_{\text{in}}$, we sort the entries of $|\left[\bar{\mathbf{y}}_{k}\right]_{n}|$ in an ascending order, and obtain a new vector $|\left[\bar{\mathbf{y}}_{k}\right]_{n}^{\uparrow}|$. Subsequently, $P_{\text{out}}$ and $P_{\text{in}}$ are estimated by
	\begin{align}		
		\setlength\abovedisplayskip{4pt}		
		\setlength\belowdisplayskip{4pt}
		\hat{P}_{\text {out }} & =\frac{1}{N_{\alpha}} \sum_{n=1}^{N_{\alpha}}|\left[\bar{\mathbf{y}}_{k}\right]_{n}^{\uparrow}|, \\
		\hat{P}_{\text {in }} & =\frac{1}{N_{\alpha}} \sum_{n=N-N_{\alpha}+1}^{N}|\left[\bar{\mathbf{y}}_{k}\right]_{n}^{\uparrow}|,
	\end{align}
	where $1\leq N_{\alpha} \ll N$ is an integer. Based on the estimated $\hat{P}_{\text{out}}$ and $\hat{P}_{\text{in}}$, we set the scaling factor $\alpha$ as
	\begin{align}	
		\setlength\abovedisplayskip{4pt}		
		\setlength\belowdisplayskip{4pt}	
		\alpha = \frac{\hat{P}_{\text{out}}+\hat{P}_{\text{in}}}{2}.
	\end{align}
	As a result, the estimated VR of user $k$ is denoted as 
	\begin{align} \label{Phi}	
		\setlength\abovedisplayskip{4pt}		
		\setlength\belowdisplayskip{4pt}	
		\hat{\boldsymbol{\Phi}}_{k}=\left\{\hat{\bar{n}}_{k}, \hat{\bar{n}}_{k}+1, \ldots, \underline{\hat{n}}_{k}-1, \underline{\hat{n}}_{k}\right\} .
	\end{align}
	
	Then, with given VR, we perform the 3D localization to facilitate the channel estimation. Let $\mathbf{X}_{k}=\left[\mathbf{x}((k-1)\tau+1), \ldots, \mathbf{x}(k\tau)\right] \in \mathbb{C}^{N \times \tau}$ denote the transmitted signals during $t \in \tilde{\mathcal T}$. Then, the received echo signal matrix at the AP from ER $k$ is denoted as
	\begin{equation}\label{receive signal}				
		\mathbf{Y}_{k}= b_{k} \mathbf{h}_{k} \mathbf{h}_{k}  ^{T}  \mathbf{X}_{k} + \mathbf{Z}_{k},
	\end{equation}	
	where $\mathbf{Y}_{k}=\left[\mathbf{y}((k-1)\tau+1), \ldots, \mathbf{y}(k\tau)\right] \in \mathbb{C}^{N \times \tau}$ and $\mathbf{Z}_{k}=\left[\mathbf{z}((k-1)\tau+1), \ldots, \mathbf{z}(k\tau)\right] \in \mathbb{C}^{N \times \tau}$ denote the echo signal matrix and the noise matrix after collecting $\tau$ consecutive symbols, respectively. Based on the received echo signal $\mathbf{Y}_{k}$ in \eqref{receive signal}, the AP estimates  position $\mathbf{l}_{k}$ of ER $k$ in the current block of interest by practical near-field localization algorithms, such as three-dimensional approximate cyclic optimization
	(3D-ACO) estimator in \cite{14}. Let $\hat{\mathbf{l}}_{k}=\{\hat{x}_{k}, \hat{y}_{k}, \hat{z}_{k}\}$ denote the estimated position of ER $k$. Then, we use estimated VR $\hat{\boldsymbol{\Phi}}_{k}$ and position $\hat{\mathbf{l}}_{k}$ for constructing the corresponding CSI of ER $k$ for energy transmission.
		
	\subsection{Energy Transmission Stage}
	Next, we consider the energy transmission in the second stage. Let $\mathbf{x}(t) \in \mathbb{C}^{N \times 1}$ denote the transmitted signal for energy transmission at symbol $t\in\{K\tau+1,\ldots, T\}$, and $\mathbf{R}_{x}=\mathbb{E}\left\{\mathbf{x}(t) \mathbf{x}^{H}(t)\right\} \succeq \boldsymbol{0}$ denote the transmit energy covariance matrix. As the energy from all beams can be
	harvested at each ER, the received RF power at ER $k$ is expressed as \footnote{The ERs that do not backscatter signals in the first stage can also harvest some energy. However, we ignore this part and focus on the harvested energy in the energy transmission stage, whcih we deem more significant due to the omnidirectional signal transmission in the first stage and the typically longer duration of the second stage.} 
	\begin{equation}		
		\begin{aligned}
			\setlength\abovedisplayskip{4pt}		
			\setlength\belowdisplayskip{4pt}
			P_{k}(\mathbf{h}_{k},\mathbf{R}_{x}) &= \left|\mathbf{h}_{k}^{H} \mathbf{x}(t)\right|^{2}.  
		\end{aligned}
	\end{equation} 	
	
	To design the transmit energy covariance matrix $\mathbf{R}_{x}$ for charging multiple ERs, we require the CSI from the AP to each ER $k$, denoted as $\mathbf{h}_{k}$. This CSI can be constructed based on the estimated VR $\hat{\boldsymbol{\Phi}}_{k}$ and position $\hat{\mathbf{l}}_{k}$ of ER $k$ in the first stage as $\hat{\mathbf{h}}_{k}=\mathbf{a}(\hat{\mathbf{l}}_{k})\odot \mathbf{g}\left(\hat{\boldsymbol{\Phi}}_{k}\right)$.
	Subsequently, the AP utilizes the constructed $\hat{\mathbf{h}}_k$ to optimize $\mathbf{R}_{x}$ for improving the estimated harvested power  $P_{k}(\hat{\mathbf{h}}_{k},\mathbf{R}_{x})$.	
	\section{Proposed Sensing Duration and Transmit Energy Beamformers Design}
	In this section, we propose an approach to design the sensing duration in the first stage and optimize the energy beamformers in the second stage. Specifically, in the first stage, the sensing duration is properly designed based on the ERs' estimation in the previous block to ensure a predetermined estimation accuracy threshold. In the second stage, the energy beamformers are designed based on the constructed CSI $\hat{\mathbf{h}}_k$ to maximum the weighted harvested power among all ERs.
	
	\vspace{-0.1cm}
	\subsection{Sensing Duration Allocation in the First Stage} 	
	In this stage,  we aims to allocate a specific duration $\tau$ to perform VR identification and position estimation. Note that allocating more duration for this stage can improve the estimation accuracy, but resulting in less duration for energy beam focusing in the second stage. Therefore, our objective is to minimize the sensing duration while guaranteeing a given requirement for estimation accuracy. In particular, we use the CRB as the performance metric for estimation accuracy, which is the performance lower bound for any unbiased estimators \cite{14}. In \eqref{receive signal}, the unknown channel parameters consist of the 3D
	position $\mathbf{l}_{k}$ and the reflection coefficient $b_{k}$. Here, we focus on estimating position $\mathbf{l}_{k}$ for one particular ER $k$. For simplicity, we omit the subscript $k$ in the following.
	
	Furthermore, we introduce the unknown parameter vector $\boldsymbol{\theta}$ $\in \mathbb{R}^{5}$ as
	\begin{equation}\label{theta}
		\setlength\abovedisplayskip{4pt}		
		\setlength\belowdisplayskip{4pt}
		\boldsymbol{\theta}=\left[x, y, z, b_{\mathrm{R}}, b_{\mathrm{I}}\right]^{T},
	\end{equation}
	where $b_{\mathrm{R}}$
	and $b_{\mathrm{I}}$
	denote the real and imaginary parts of $b$,
	respectively. The sample covariance
	matrix in this stage is given by $\mathbf{S}_{x}=\frac{1}{\tau}\mathbf{X}\mathbf{X}^{H}$. The Fisher information matrix (FIM) $\mathbf{F} \in \mathbb{R}^{5 \times 5}$ of the parameter vector $\boldsymbol{\theta}$ is given by \cite{14}
	\begin{equation}
		\setlength\abovedisplayskip{4pt}		
		\setlength\belowdisplayskip{4pt}
		\frac{2}{\sigma_{r}^{2}}\left[\begin{array}{ccccc}
			\mathfrak{R}\left(\mathbf{F}_{xx}\right) & \!\!\!\!\mathfrak{R}\left(\mathbf{F}_{xy}\right) & \!\!\!\!\mathfrak{R}\left(\mathbf{F}_{xz}\right) &\!\!\!\! \mathfrak{R}\left(\mathbf{F}_{xb}\right) &\!\!\!\! -\mathfrak{I}\left(\mathbf{F}_{xb}\right) \\
			\mathfrak{R}\left(\mathbf{F}_{xy}^{T}\right) &\!\!\!\! \mathfrak{R}\left(\mathbf{F}_{yy}\right) &\!\!\!\! \mathfrak{R}\left(\mathbf{F}_{yz}\right) &\!\!\!\!\mathfrak{R}\left(\mathbf{F}_{yb}\right) &\!\!\!\! -\mathfrak{I}\left(\mathbf{F}_{yb}\right) \\
			\mathfrak{R}\left(\mathbf{F}_{xz}^{T}\right) &\!\!\!\! \mathfrak{R}\left(\mathbf{F}_{yz}^{T}\right) &\!\!\!\! \mathfrak{R}\left(\mathbf{F}_{zz}\right) &\!\!\!\! \mathfrak{R}\left(\mathbf{F}_{zb}\right) &\!\!\!\! -\mathfrak{I}\left(\mathbf{F}_{zb}\right) \\
			\mathfrak{R}\left(\mathbf{F}_{xb}^{T}\right) &\!\!\!\! \mathfrak{R}\left(\mathbf{F}_{yb}^{T}\right) &\!\!\!\! \mathfrak{R}\left(\mathbf{F}_{zb}^{T}\right) &\!\!\!\! \mathfrak{R}\left(\mathbf{F}_{bb}\right) &\!\!\!\! -\mathfrak{I}\left(\mathbf{F}_{bb}\right) \\
			-\mathfrak{I}\left(\mathbf{F}_{xb}^{T}\right) &\!\!\!\! -\mathfrak{I}\left(\mathbf{F}_{yb}^{T}\right) &\!\!\!\! -\mathfrak{I}\left(\mathbf{F}_{zb}^{T}\right) &\!\!\!\! -\mathfrak{I}\left(\mathbf{F}_{bb}^{T}\right) &\!\!\!\! \mathfrak{R}\left(\mathbf{F}_{bb}\right)
		\end{array}\right],
	\end{equation}
	where $\mathbf{F}_{xx}$, $\mathbf{F}_{yy}$, and $\mathbf{F}_{zz}$ are given by
	\begin{equation}
		\begin{aligned}
			\setlength\abovedisplayskip{4pt}		
			\setlength\belowdisplayskip{4pt}
			\mathbf{F}_{uu} & =\tau \left|b\right| \left(\dot{\mathbf{h}}_{u}^{H}  \dot{\mathbf{h}}_{u}\right)  \left(  \mathbf{h}^{H} \mathbf{S}_{x}^{*} \mathbf{h}  \right) 
			 +\tau \left|b\right|\left(\dot{\mathbf{h}}_{u}^{H}  \mathbf{h}\right)  \left(  \mathbf{h}^{H} \mathbf{S}_{x}^{*} \dot{\mathbf{h}}_{u}  \right) \\
			& +\tau \left|b\right|\left(\mathbf{h}^{H}  \dot{\mathbf{h}}_{u}\right)  \left(  \dot{\mathbf{h}}_{u}^{H} \mathbf{S}_{x}^{*} \mathbf{h}  \right)  
			  +\tau \left|b\right|\left(\mathbf{h}^{H}  \mathbf{h}\right)  \left(  \dot{\mathbf{h}}_{u}^{H} \mathbf{S}_{x}^{*} \dot{\mathbf{h}}_{u}  \right),
		\end{aligned}
	\end{equation}
	with $\tilde{x} \in\{\tilde{x}, \tilde{y}, \tilde{z}\}$,
	$\mathbf{F}_{xy}$, $\mathbf{F}_{xz}$, and $\mathbf{F}_{yz}$ are given by
	\begin{equation}		
		\begin{aligned}
			\setlength\abovedisplayskip{4pt}		
			\setlength\belowdisplayskip{4pt}
			\mathbf{F}_{uv}&=\tau \left|b\right|\left(\dot{\mathbf{h}}_{u}^{H}  \dot{\mathbf{h}}_{v}\right)  \left(  \mathbf{h}^{H} \mathbf{S}_{x}^{*} \mathbf{h}  \right) 
			+\tau \left|b\right|\left(\dot{\mathbf{h}}_{u}^{H}  \mathbf{h}\right) \left(  \mathbf{h}^{H} \mathbf{S}_{x}^{*} \dot{\mathbf{h}}_{v}  \right) \\
			&+\tau \left|b\right|\left(\mathbf{h}^{H}  \dot{\mathbf{h}}_{v}\right) \left(  \dot{\mathbf{h}}_{u}^{H} \mathbf{S}_{x}^{*} \mathbf{h}  \right) 
			+\tau \left|b\right|\left(\mathbf{h}^{H}  \mathbf{h}\right)  \left(  \dot{\mathbf{h}}_{u}^{H} \mathbf{S}_{x}^{*} \dot{\mathbf{h}}_{v}  \right), 
		\end{aligned}
	\end{equation}
	with $uv \in\{\tilde{x}\tilde{y}, \tilde{x}\tilde{z}, \tilde{y}\tilde{z}\}$,  $\mathbf{F}_{bb}$, $\mathbf{F}_{xb}$, $\mathbf{F}_{yb}$, and $\mathbf{F}_{zb}$ are given by
	\begin{align}	
		\setlength\abovedisplayskip{4pt}		
		\setlength\belowdisplayskip{4pt}	
		\mathbf{F}_{bb}  &= \tau \left(\mathbf{h}^{H}  \mathbf{h}\right)  \left(\mathbf{h}^{H} \mathbf{S}_{x}^{*} \mathbf{h}\right),\\ 
		\mathbf{F}_{u b} & =\tau \left(\dot{\mathbf{h}}_{u}^{H}  \mathbf{h}\right)  \left( b^{*} \mathbf{h}^{H} \mathbf{S}_{x}^{*} \mathbf{h}\right) + \tau\left(\mathbf{h}^{H}  \mathbf{h}\right)  \left( b^{*} \dot{\mathbf{h}}_{u}^{H} \mathbf{S}_{x}^{*} \mathbf{h}\right),
	\end{align}
	respectively. Here, we have
	\begin{align}	
		\setlength\abovedisplayskip{4pt}		
		\setlength\belowdisplayskip{4pt}
		\dot{\mathbf{h}}_{u}&=\frac{\partial \mathbf{a}\left(\mathbf{l}\right)}{\partial u} \odot  \mathbf{g}\left(\boldsymbol{\Phi}\right),\\
		\left(\frac{\partial \mathbf{a}\left(\mathbf{l}\right)}{\partial u}\right)_{n}&=\mathbf{a}_{n}\left(\mathbf{l}\right)\left(\frac{u_{n}^{t}-u}{\left\|\mathbf{l}^{t}_{n}-\mathbf{l}\right\|^{2}}+\mathrm{j} \frac{2 \pi}{\lambda} \frac{u_{n}^{t}-u}{\left\|\mathbf{l}^{t}_{n}-\mathbf{l}\right\|}\right),
	\end{align}
	where $\mathbf{a}_{n}(\mathbf{l})$ denotes the $n \text{-th}$ element of $\mathbf{a}({\mathbf{l}})$. Consequently, the CRB  matrix $\mathbf{C}$ for estimating position $\boldsymbol{\theta}$ is calculated as $\mathbf{C}=\mathbf{F}^{-1}$. Referring to \eqref{theta}, the total CRB for estimating $\mathbf{l}$ under a given duration $\tau$ is derived as 
	\begin{equation}\label{CRB_k}
		\setlength\abovedisplayskip{4pt}		
		\setlength\belowdisplayskip{4pt}
		\text{CRB}(\tau,x,y,z)=\mathbf{C}[1,1]+\mathbf{C}[2,2]+\mathbf{C}[3,3],
	\end{equation}
	where $\mathbf{C}[i,i]$ denotes the $i \text{-th}$ diagonal entry of CRB matrix $\mathbf{C}$. Let  $\bar{\mathbf{l}} _{k}=\left(x_{k}+\Delta x_{k}, y_{k}+\Delta y_{k}, z_{k}+\Delta z_{k}\right)$ denote the estimated position in the previous
	block, where $\Delta x_{k}$, $\Delta y_{k}$ and $\Delta z_{k}$ denote the corresponding localization errors. It is assumed that $\Delta x_{k}$, $\Delta y_{k}$ and $\Delta z_{k}$ are random variables that are bounded, i.e., $|\Delta x_{k}| \leq D_{x}$,  $|\Delta y_{k}| \leq D_{y}$ and $|\Delta z_{k}| \leq D_{z}$ with $D_{x}$, $D_{y}$ and $D_{z}$ denoting the corresponding error bounds. Based on the CRB in \eqref{CRB_k}, we set duration $\tau^{\star}$ as the minimum value such that the estimation CRBs for all ERs do not exceed estimation accuracy threshold $\Gamma$, i.e., $\mathop{\max}\limits_{k\in\mathcal K}\text{CRB}(\tau^{\star},\bar{x}_{k}, \bar{y}_{k}, \bar{z}_{k})\leq\Gamma$.	
	
	\subsection{Energy Beamfroming Design in the Second Stage}
	In this stage, we consider the energy transmission during the remaining duration of $T-K\tau$. In particular, considering that different IoT devices may have varying energy requirements, the transmit energy covariance matrix $\mathbf{R}_{x}$ is designed to maximize the weighted harvested RF power based on constructed CSI $\hat{\mathbf{h}}_{k}$. The corresponding problem is formulated as 
	\begin{subequations}		  
		\begin{align}
			\setlength\abovedisplayskip{4pt}		
			\setlength\belowdisplayskip{4pt}
			(\text{P4}):	\max _{\mathbf{R}_{x}\succeq 0} 
			&\quad \sum_{k=1}^{K}  \beta_{k} \hat{\mathbf{h}}_{k}^{H} \mathbf{R}_{x} \hat{\mathbf{h}}_{k}\\ 
			\mathrm{s.t.}
			& \quad\operatorname{tr}\left(\mathbf{R}_{x}\right)\leq P_{\text{max}}\label{C23}.
		\end{align}
	\end{subequations}
	where $\beta_{k} \geq 0$ are predefined energy weight of ER $k$ that are application-specific, and a larger value of $\beta_{k}$ signifies a higher priority on transmitting energy to ER $k$ in comparison to other ERs. Problem (P4) is a convex problem that can be solved by using CVX \cite{16}. With optimal solution $\mathbf{R}_{x}^{\star}$ obtained, the corresponding average harvested power of ER $k$ is given by	
	$\frac{(T-K\tau) }{T}\mathbf{h}_{k}^{H} \mathbf{R}_{x}^{\star} \mathbf{h}_{k}$, where $\mathbf{h}_{k}$ is the exact channel between the AP and ER $k$.
	
	\begin{remark}\label{remark}
	Note that in our proposed design, the time allocation between the two stages should be suitably design for maximizing the ultimate harvested energy. According to CRB mentioned in \eqref{CRB_k},  the sensing accuracy is directly proportional to $\tau$. In particular, increasing the time duration allocated for  radar sensing in the first stage achieves more accurate channel parameters estimation but reduces the time available for energy transmission in the second stage. This effect will be illustrated in the numerical results in the Section IV.
	\end{remark} 
	\vspace{-0.15cm}
	\section{Numerical Results}	
		\vspace{-0.15cm}
	This section provides numerical results to validate the effectiveness of our proposed sensing-assisted energy beam focusing design. In the simulation, we set the UPA with total antenna number $N = 16 \times 16 = 256$. We set the noise power as $\sigma_{r}^{2}=-120$ dBm.  The spacing
	between adjacent antennas is half-wavelength with a carrier frequency of 28 GHz. We consider that there are $K=2$ ERs in the system, in which their 3D positions are given by $\bar{\mathbf{l}}_{1}=[1,2,3]$ m and $\bar{\mathbf{l}}_{2}=[1.5,3,4.5]$ m, respectively. We set all error bounds as $D_{x}=D_{y}=D_{z}=0.15$ m. Their energy weights are set as $\beta_{1}=0.1$ and $\beta_{2}=0.9$ unless otherwise specified. The VR are randomly configured as a sub-array, with the proportional factor of VR set to $\eta=\frac{1}{4}$. We set the integer $N_{\alpha}$ as 32. Furthermore, we assume that each transmission block consists of $T=200$ symbols.
	\begin{figure}
		\setlength{\abovecaptionskip}{-16pt}
		\setlength{\belowcaptionskip}{-16pt}
		\centering
			\centering
			\includegraphics[width=0.4\textwidth]{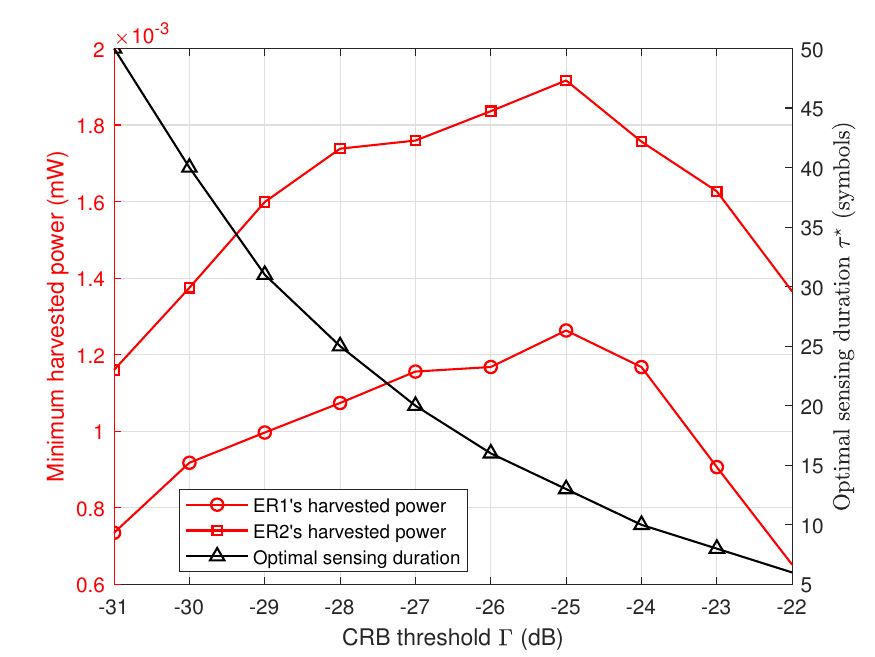}
		\DeclareGraphicsExtensions.
		\captionsetup{font={scriptsize},skip=3pt}
		\caption{Harvested power and optimized sensing duration versus the	CRB threshold $\Gamma$.}
		\label{figure3}		
	\end{figure}	
	
	Fig.~\ref{figure3} shows the average harvested powers of ERs (i.e., energy normalized by the block duration $T$) and the sum of CRB versus the sensing duration $\tau$, with the maximum transmit power set to $P_{\text{max}}=30$ dBm. It is observed that there exists an optimal value of $\Gamma$ that enables both ERs to achieve maximum harvested power. When $\Gamma$ is below this threshold, the harvested power of both ERs increases as $\Gamma$ increases, because the AP requires less sensing duration in the first stage to guarantee estimation accuracy, allowing for more remaining duration in the second stage to execute energy beam focusing. As $\Gamma$ continues to increase, the harvested power of both ERs are seen to decrease because their estimated parameters become less accurate. This shows an trade-off between sensing accuracy and energy transmission.
	
	In the following, we compare the performance of our proposed design  with the following benchmark schemes. 
	\begin{itemize}	
		\item \textbf {Perfect CSI: }The AP designs energy beam focusing by solving probliem (P4) under the assumption of perfect knowledge of the CSI  $\{\mathbf{h}_{k}\}$, with the whole transmission block allocated for energy transmission.
		\item \textbf {Isotropic transmission: }The AP uses the identity transmit covariance matrix $\mathbf{R}_{x}=\frac{P_{\text{max}}}{N}\mathbf{I}$ for energy transmission in the whole transmission block. 
		\item \textbf {Equal time allocation: }The two stages are of equal duration for radar sensing and energy transmission, with the  beamformers designed based on those in Section III. 
		\item \textbf {Design without VR identification: }The AP designs the energy beamformers based on the two-stage protocol, but it does not identify the VRs of ERs in the first stage.
	\end{itemize}
	\begin{figure}[ht]
		\vspace{-0.2cm}
		\setlength{\abovecaptionskip}{-8pt}
		\setlength{\belowcaptionskip}{-8pt}
		\begin{minipage}[t]{0.48\columnwidth}
			\centering
			\includegraphics[width=1\textwidth]{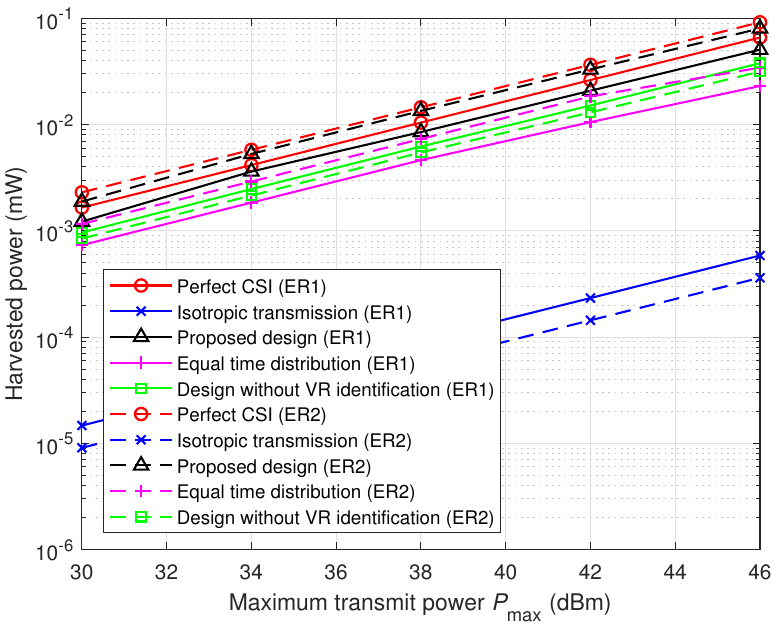}
			\DeclareGraphicsExtensions.
			\captionsetup{font={scriptsize},skip=1pt}			
			\caption{ERs' harvested powers versus the maximum transmit power $P_{\text{max}}$.}
			\label{figure4}
		\end{minipage}\hspace{+0.15cm}
		\begin{minipage}[t]{0.48\columnwidth}
			\centering
			\includegraphics[width=1\textwidth]{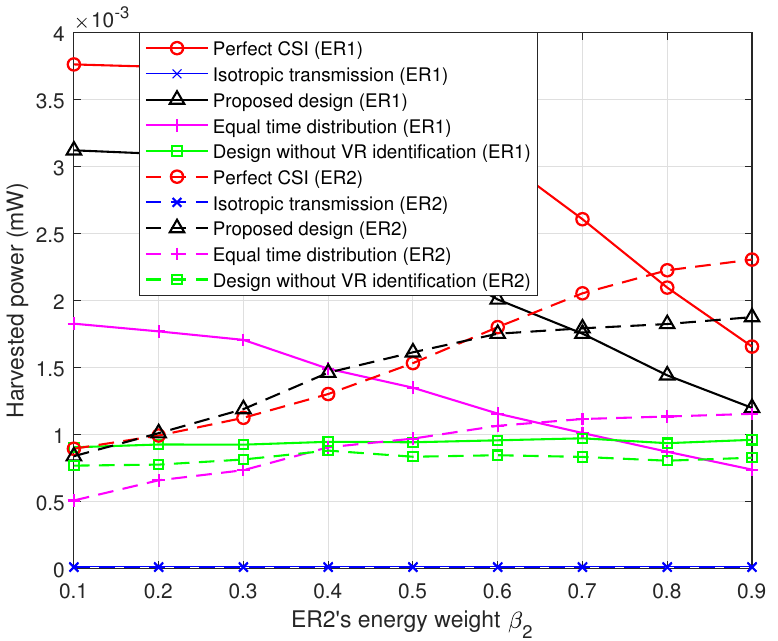}
			\DeclareGraphicsExtensions.
			\captionsetup{font={scriptsize},skip=1pt}
			\caption{ERs' harvested powers versus the ER2's energy weight $\beta_{2}$.}
			\label{figure5}
		\end{minipage}
	\end{figure}
	
	Fig. \ref{figure4} shows harvested powers of ERs versus the maximum transmit power. It is shown that our proposed design achieves performance close to the upper bound with perfect CSI, verifying the effectiveness of our scheme. Additionally, the results indicates that our design achieves superior performance compared to benchmark schemes due to the energy beam focusing capability and the efficient time allocation. Furthermore, despite ER1 and ER2 are located in the same direction and ER2 is further away from the AP, the harvested power by ER2 in our proposed design is greater than that harvested by ER1. This is because ER2's energy weight is set to be larger than that assigned to ER1, allowing the beamformers to be directed to the specific location of ER2 in our design, rather than just the general direction in which they are located.	
	
	Fig. \ref{figure5} shows harvested powers of ERs versus the ER2's energy weight $\beta_{2}$ with the maximum transmit power set to $P_{\text{max}}=30$ dBm. It is observed that in the proposed design, as the ER2's energy weight increases, the energy received by ER2 gradually increases, while the energy received by ER1 decreases gradually. Therefore, we can design appropriate energy weights based on the actual charging needs of the ERs. 
	\vspace{-0.2cm}
	\section{Conclusion}
	\vspace{-0.15cm}
	In this paper, we proposed a sensing-assisted near-field energy beam focusing approach for wirelessly charging multiple ERs without requiring explicit channel training and feedback. We presented a two-stage protocol, where the AP transmits radar sensing signals to identify the VR and estimate 3D positions of each ER in the first stage  for constructing CSI, and subsequently, the AP designs energy beam focusing based on the constructed CSI in the second stage to maximize weighted harvested energy across all ERs. In our approach, we designed proper time allocation between the two stages to optimize the energy harvesting performance. Numerical results showed that the proposed design achieves performance close to the upper bound with perfect VR and CSI and superior performance to benchmark schemes. This paper offers perspectives on future near-field ISCAP networks \cite{12}.
	

\vspace{-0.15cm}
	\begin{thebibliography}{00}	
		\vspace{-0.15cm}	
		\bibitem{1}	
		X. You et al., ``Towards 6G wireless communication networks: Vision, enabling technologies, and new paradigm shifts,'' \textit{Sci. China Inf. Sci.}, vol. 64, no. 1, pp. 1-74, Jan. 2021.	
		
		\bibitem{2}
		Y. Zeng, B. Clerckx, and R. Zhang, ``Communications and signals design for wireless power transmission,'' \textit{
			IEEE Trans. Commun.}, vol. 65, no. 5, pp. 2264-2290, May. 2017.		
		
		\bibitem{4}
		H. Zhang, N. Shlezinger, F. Guidi, D. Dardari, M. F. Imani, and Y. C. Eldar, ``Beam focusing for near-field multiuser MIMO communications,'' \textit{IEEE Trans. Wireless Commun.}, vol. 21, no. 9, pp. 7476-7490, Sept.	2022.
		
		\bibitem{11}
		H. Hua, J. Xu, and R. Zhang, ``Near-field integrated sensing and communication with extremely large-scale antenna array,'' 2024. [Online] Available: https://arxiv.org/abs/2407.17237
		
		\bibitem{12}
		Z. Ren, S. Zhang, X. Li, L. Qiu, J. Xu, and D. W. K. Ng, ``Secure communications in near-filed ISCAP systems with extremely large-scale antenna arrays,'' in \textit{IEEE Int. Symp. Wirel. Commun. Syst.} (\textit{ISWCS}), 2024.	
		
		\bibitem{6}
		M. Cui, Z. Wu, Y. Lu, X. Wei, and L. Dai, ``Near-field MIMO
		communications for 6G: Fundamentals, challenges, potentials, and future
		directions,'' \textit{IEEE Commun. Mag.}, vol. 61, no. 1, pp. 40-46, Jan. 2023.
		
		\bibitem{8}
		M. Cui and L. Dai, ``Channel estimation for extremely large-scale
		MIMO: Far-field or near-field?'' \textit{IEEE Trans. Commun.}, vol. 70, no.
		4, pp. 2663-2677, Apr. 2022.
		
		\bibitem{9}
		J. Zhang, J. Zhang, Y. Han, J. Wang, and S. Jin, ``Average spectral efficiency for TDD-based non-stationary XL-MIMO with VR estimation,'' in \textit{Proc. IEEE Wireless Commun. Signal Process.} (\textit{WCSP}), 2022, pp. 973-977.
		
		\bibitem{18}
		F. Liu, Y. Cui, C. Masouros, J. Xu, T. X. Han, Y. C. Eldar, and S. Buzzi,
		``Integrated sensing and communications: Toward dual-functional wireless networks for 6G and beyond,'' \textit{IEEE J. Sel. Areas Commun.}, vol. 40,
		no. 6, pp. 1728–1767, Jun. 2022.
		
		\bibitem{10}
		Y. Chen, H. Hua, J. Xu, and D. W. K. Ng, ``ISAC meets SWIPT:
		Multi-functional wireless systems integrating sensing, communication,
		and powering,'' \textit{IEEE Trans. Wireless Commun.}, vol. 23, no. 8, pp. 8264-8280, Aug. 2024.
		
		\bibitem{17}
		Y. Chen, Z. Ren, J. Xu, Y. Zeng, D. W. K. Ng, and S. Cui, ``Integrated sensing, communication, and powering (ISCAP): Towards multi-functional 6G wireless networks,'' 2024. [Online] Available: https://arxiv.org/abs/2401.03516
		
		\bibitem{19}
		F. Liu, W. Yuan, C. Masouros, and J. Yuan, ``Radar-assisted predictive beamforming for vehicular links: Communication served by sensing,'' \textit{IEEE Trans. Wireless Commun.}, vol. 19, no. 11, pp. 7704-7719, Nov. 2020.
		
		\bibitem{21}
		Z. Ren, L. Qiu, J. Xu, and D. W. K. Ng, ``Sensing-assisted sparse channel recovery for massive antenna systems,'' \textit{IEEE Trans. Veh. Technol.}, early access, doi: 10.1109/TVT.2024.3422663. 
		
		\bibitem{13}
		L. Zhang, Y. Fang, Z. Ren, L. Qiu, and J. Xu, ``Training-free energy beam focusing assisted by wireless sensing,'' in \textit{Proc. IEEE Wireless Commun. Networking Conf.}   (\textit{WCNC}), 2024, pp. 1-6.
		
		\bibitem{14}
		H. Hua, J. Xu, and Y. C. Eldar, ``Near-field 3D localization via MIMO radar: Cramer-Rao bound analysis and estimator design,''  
		\textit{IEEE Trans. Signal Process.}, early access, 2024, doi: 10.1109/TSP.2024.3441815.
		
		\bibitem{15}
		Y. Han, S. Jin, C. -K. Wen, and T. Q. S. Quek, ``Localization and channel reconstruction for extra large RIS-assisted massive MIMO systems,'' \textit{IEEE J. Sel. Top. Signal Process.}, vol. 16, no. 5, pp. 1011-1025, Aug. 2022.				
		
		\bibitem{16}
		M. Grant and S. Boyd, ``CVX: Matlab software for disciplined
		convex programming, version 2.1,'' 2014. [Online]. Available:
		http://cvxr.com/cvx
		
			
		
		
		
		
	\end{thebibliography}
\end{document}